%% file: main.tex
\documentclass[10pt,conference]{IEEEtran}
\IEEEoverridecommandlockouts
\usepackage{cite}
\usepackage{amsmath,amssymb,amsfonts}
\usepackage{algorithmic}
\usepackage{graphicx}
\usepackage{textcomp}
\usepackage{xcolor}
\usepackage{booktabs}
\def\BibTeX{{\rm B\kern-.05em{\sc i\kern-.025em b}\kern-.08em
    T\kern-.1667em\lower.7ex\hbox{E}\kern-.125emX}}

\usepackage{braket}
\usepackage{mathtools}
\usepackage{hyperref}
\usepackage[draft, commentmarkup=todo,
   todonotes={textsize=tiny, textwidth=0.83in}]{changes}
\definechangesauthor[name={Bao Bach}, color=teal]{bb}
\definechangesauthor[name={Kien Nguyen}, color=blue]{kn}
\definechangesauthor[name={Ilya Safro}, color=red]{is}

\begin{document}

\title{Cross-Problem Parameter Transfer in Quantum Approximate Optimization Algorithm:\\ A Machine Learning Approach}

\author{\IEEEauthorblockN{Kien X. Nguyen}
\IEEEauthorblockA{
\textit{Computer and Information Sciences} \\
\textit{University of Delaware}\\
Newark, Delaware \\
kxnguyen@udel.edu}
\and
\IEEEauthorblockN{Bao Bach}
\IEEEauthorblockA{
\textit{Computer and Information Sciences} \\
\textit{Quantum Science and Engineering} \\
\textit{University of Delaware}\\
Newark, Delaware \\
baobach@udel.edu}
\and
\IEEEauthorblockN{Ilya Safro}
\IEEEauthorblockA{\textit{Computer and Information Sciences} \\
\textit{Physics and Astronomy} \\
\textit{University of Delaware}\\
Newark, Delaware \\
isafro@udel.edu}
}

\maketitle

\input{sec/abstract}


\input{sec/intro}
\input{sec/related_work}
\input{sec/background}
\input{sec/method}
\input{sec/experiment}
\input{sec/conclusion}

\clearpage
\bibliographystyle{IEEEtran}
\bibliography{main}

\end{document}

%% file: sec/abstract.tex
\begin{abstract}
Quantum Approximate Optimization Algorithm (QAOA) is one of the most promising candidates to achieve the quantum advantage in solving combinatorial optimization problems.
The process of finding a good set of variational parameters in the QAOA circuit has proven to be challenging due to multiple factors, such as barren plateaus.
As a result, there is growing interest in exploiting \textit{parameter transferability}, where parameter sets optimized for one problem instance are transferred to another that could be more complex either to estimate the solution or to serve as a warm start for further optimization. But can we transfer parameters from one class of problems to another? Leveraging parameter sets learned from a well-studied class of problems could help navigate the less studied one, reducing optimization overhead and mitigating performance pitfalls.
In this paper, we study whether pretrained QAOA parameters of MaxCut can be used as is or to warm start the Maximum Independent Set (MIS) circuits.
Specifically, we design machine learning models to find good donor candidates optimized on MaxCut and apply their parameters to MIS acceptors.
Our experimental results show that such parameter transfer can significantly reduce the number of optimization iterations required while achieving comparable approximation ratios.\\
\noindent {\bf Reproducibility: } Our results and code are available at: \url{https://github.com/Nyquixt/Cross-Problem-PT-QAOA}.
\end{abstract}


%% file: sec/intro.tex
\begin{figure*}
    \centering
    \includegraphics[width=0.8\linewidth]{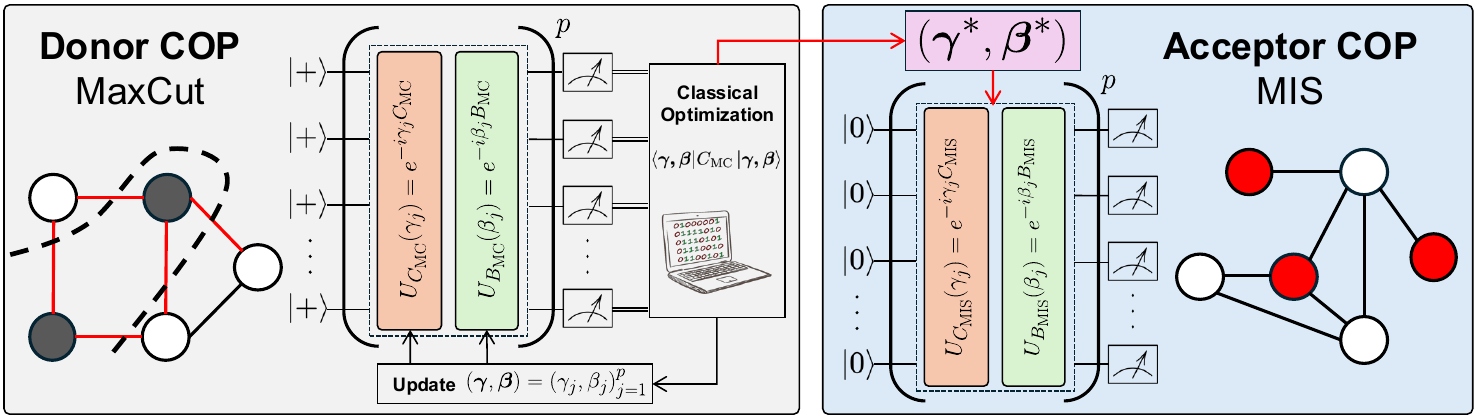}
    \caption{Overview of the parameter transfer procedure. First, QAOA is run on a quantum circuit to optimize for a \textit{donor} COP (i.e. MaxCut) and retrieve the optimal parameter set $(\boldsymbol{\gamma}^*, \boldsymbol{\beta}^*)$. Next, those parameters are directly applied to the \textit{acceptor} COP (i.e. MIS) without additional tuning.}
    \label{fig:title}
\end{figure*}

\section{Introduction}
The class of quantum approximate optimization algorithm (QAOA) \cite{farhi2014quantumapproximateoptimizationalgorithm} and its variants is one of the most promising candidates to achieve the quantum advantage in the mathematical optimization domain.
Its adaptable structure and relatively shallow circuit depth make it well-suited for near-term quantum devices, unlocking a broad range of potential applications, such as finance~\cite{herman2023quantum}, biology~\cite{outeiral2021prospects}, and scientific computing~\cite{lin2022lecture}.

QAOA is a variational algorithm with a direct application to solve combinatorial optimization problems (COPs).
This hybrid quantum-classical algorithm consists of a parameterized quantum circuit whose parameters are iteratively updated by a classical optimizer.
Despite recent progress in the field, effectively tuning these parameters remains a significant challenge, as the optimization landscape can exhibit barren plateaus when the depth of the circuit or the number of qubits increases~\cite{mcclean2018barren, anschuetz2022quantum, wang2021noise, kulshrestha2204beinit}.
Furthermore, scaling QAOA to larger problem instances often entails rapidly increasing circuit depths, which not only complicates parameter optimization but also exacerbates hardware-induced errors on near-term quantum devices~\cite{zhou2020quantum}.

Such challenges have prompted research on parametric initialization methods, such as linear ramp~\cite{zhou2020quantum,montanez2024towards} or multistart optimization~\cite{shaydulin2019multistart}.
Another emerging strategy to overcome the difficulty of finding good parameters is \textit{parameter transfer}, wherein parameter sets that have already been optimized for one QAOA instance are reused on another~\cite{montanez2402transfer,brandao2018fixed,galda2021transferability,falla2024graph}.
For example, Brand\~ao et al.~\cite{brandao2018fixed} demonstrated that QAOA parameters optimized for the MaxCut problem on a specific 3-regular graph remain nearly optimal across all other 3-regular graphs.
Galda et al.~\cite{galda2021transferability} identified different properties of random graphs as a proxy for parameter transferability.
Sureshbabu et al.~\cite{sureshbabu2024parameter} introduced a straightforward rescaling method to transfer parameters from unweighted to weighted graphs for MaxCut.

While these efforts illustrate the effectiveness of parameter transfer within the same COP, an important next step is to explore whether such transferability extends to fundamentally different COPs~\cite{montanez2402transfer}. We explore the following question: given a sufficiently large set of QAOA preoptimized circuits (donors) for one COP, can we build a low resource transferability method for circuits (acceptors) of another COP. 
Monta\~nez et al. note that it remains unclear why certain COPs (i.e. Bin Packing) exhibit better generalization capability than others, and their empirical studies are based on random instances.

\noindent {\bf Our contribution: }To address this open question, we extend beyond purely random instances and adopt a data-driven approach, leveraging machine learning (ML) techniques to systematically \textit{identify good donor instances} optimized on an unconstrained COP MaxCut, transferring their optimal parameters to a more challenging, constrained COP Maximum Independent Set (MIS).
In particular, we use graph low dimensional representations (embeddings) to train a neural network that predicts a \textit{transfer score}, reflecting how well the parameters from a donor graph will perform on an acceptor graph.
By encoding both graphs into low-dimensional embeddings and framing donor selection as a learning task, we capture structural relationships that drive parameter transferability and enable us to discover parameter sets that generalize effectively to new or more complex problems.

Our empirical results show that the proposed ML model is capable of identifying suitable donors that could transfer their optimized MaxCut parameters to their corresponding accepter on MIS and achieve an approximation ratio of 0.8385 without further tuning.
We also conduct further fast optimization on MIS taken the transferred MaxCut parameters as initialization and show that it only takes 10 steps to improve to an approximation ratio of 0.9500 on which is insufficient otherwise.

%% file: sec/related_work.tex
\section{Related Work}

\subsection{Parameter Transferability in QAOA}
Finding good QAOA parameters remains difficult due to various challenges, such as barren plateaus~\cite{anschuetz2022quantum,wang2021noise,kulshrestha2204beinit}.
Although deeper QAOA circuits can deliver slight improvements in the approximation ratio, they greatly increase computational overhead.
Consequently, rather than optimizing parameters from scratch for every new instance, especially for significantly larger ones, a practical strategy is to reuse parameter sets previously optimized for a ``related" instance~\cite{galda2021transferability,montanez2402transfer,falla2024graph}.
The transferred parameters from the donor instance can either be applied directly to the accepter instance, as demonstrated in this work, or be used as a warm start for further optimization.
Recent studies have begun examining the transferability of QAOA parameters across different COPs~\cite{montanez2402transfer}.
Monta\~nez et al.~\cite{montanez2402transfer} present a numerical study on transferring QAOA parameters between different COPs.
Although their results hint at the potential for cross-problem parameter transferability, the scope of their analysis remains limited to a small collection of randomly generated instances.
In comparison, we adopt a more comprehensive, data-driven approach, leveraging machine learning techniques to systematically identify good donor candidates on the simpler, unconstrained MaxCut problem, and transfer their parameters to the more challenging, constrained MIS problem.

\subsection{Machine Learning for Quantum Computing}
Broadly, the intersection of machine learning and quantum computing can be divided into two main areas: Quantum Machine Learning (QML) and Machine Learning for Quantum Computing (ML for QC).
QML focuses on harnessing quantum hardware to accelerate or enhance traditional ML tasks, offering speedups for problems such as classification, clustering, or generative modeling~\cite{biamonte2017quantum,zaman2023survey,cong2019quantum,huang2021experimental}.
In contrast, ML for QC applies classical learning techniques to improve or optimize quantum algorithms and devices, tackling challenges such as parameter tuning~\cite{verdon2019learning}, noise mitigation~\cite{xu2025physics}, or circuit architecture search~\cite{duong2022quantum}.
In this paper, we use machine learning to predict the good donor candidates for parameter transferability, which can be considered as a retrieval task~\cite{chen2022deep}.



%% file: sec/background.tex
\section{Background}
\subsection{Quantum Approximate Optimization Algorithm}

The QAOA~\cite{farhi2014quantumapproximateoptimizationalgorithm} is a hybrid quantum-classical algorithm usually applied to combinatorial optimization problems. The quantum heuristic algorithm prepares a parameterized circuit to approximate the solution to the problem. The variational loop contains $p$ layers of alternating unitaries and a classical optimizer. The optimizer's role is to find the optimal variational parameters.

For a combinatorial optimization problem with the cost function (objective)  $\mathcal{C}(x)$ and $x \in \{0, 1\}^n$, the QAOA circuit alternates between the problem unitary $U_C(\gamma)=e^{-i\gamma C}$ and the mixing unitary $U_B(\beta)=e^{-i\beta B}$ from two Hamiltonian families $C$ and $B$ for $p$ layers, parameterized by $\boldsymbol{\gamma} = \{\gamma_i\}$ and $\boldsymbol{\beta} = \{\beta_i\}$ with $1 \leq i \leq p$, and prepares the final state
\begin{equation}
    \ket{\boldsymbol{\gamma}, \boldsymbol{\beta}} = U_B(\beta_p)U_C(\gamma_p)\dots U_B(\beta_1)U_C(\gamma_1)\ket{s}^{\otimes n},
\end{equation}
\noindent where $\ket{s}$ is the initial state. The classical optimizer then aims to find the optimal $(\boldsymbol{\gamma}^*, \boldsymbol{\beta}^*)$ with respect to the expectation value
\begin{equation}
    \braket{C(\boldsymbol{\gamma}^*, \boldsymbol{\beta}^*)} \coloneq \bra{\boldsymbol{\gamma}^*, \boldsymbol{\beta}^*}C\ket{\boldsymbol{\gamma}^*, \boldsymbol{\beta}^*} = \sum_{x\in \{0,1\}^n} \mathcal{C}(x) \text{Pr}(x),
\end{equation}
\noindent where $\text{Pr}(x)$ is the probability of observing $x$ when measuring all qubits of $\ket{\boldsymbol{\gamma}, \boldsymbol{\beta}}$.
The approximation ratio is calculated as
\begin{equation}
    r = \frac{\bra{\boldsymbol{\gamma}^*, \boldsymbol{\beta}^*}C\ket{\boldsymbol{\gamma}^*, \boldsymbol{\beta}^*}}{C_{\text{opt}}},
\end{equation}
where $C_{\text{opt}}$ is the optimal solution for the problem instance. We use the Gurobi solver to find the classical optimal solution.

\subsection{Maximum Cut and Maximum Independent Set}
\noindent\textbf{Maximum Cut.}
The Maximum Cut problem (MaxCut) involves finding a cut that splits a graph into two disjoint partitions so that the weighted sum of the edges within the cut is maximized. MaxCut, an NP-complete problem, can be formulated as a quadratic unconstrained optimization problem (QUBO) by assigning a binary value $x_i$ to every vertex based on the partition it belongs to.

Given a weighted graph $G=(V,E,w)$, where $V$ is the set of vertices, $E$ denotes the set of edges and $w$ represents the edge weighting function, the MaxCut problem is defined as
\begin{equation}
    \max_{x\in\{-1,1\}^n} \sum_{(i,j)\in E} w_{ij}\frac{1 - x_ix_j}{2}.
\end{equation}
The cost Hamiltonian $C_{\text{MC}}$ can be constructed as $\sum_{(i,j)\in E} \frac{1}{2}(Z_iZ_j - \mathbb{I})$, where $Z$ is the Pauli operator $Z$. 
The mixing Hamiltonian $B_{\text{MC}}$ is constructed as $\sum_{i\in V} X_i$, where $X$ is the Pauli operator $X$. The initial quantum state for classical QAOA MaxCut is $\ket{+}^{\otimes n}$.

\vspace{5pt}
\noindent\textbf{Maximum Independent Set.}
For unweighted graph $G=(V,E)$, a set $I\subseteq V$ is an {\it independent set} if for all $i,j\in I$, $ij\not\in E$. A maximum independent set (MIS) in $G$ is an independent set of largest cardinality. Finding MIS is NP-hard.

For graph $G$, a subset $V' \subset V$ can be represented by its indicator vector $\mathbf{x} \in \{0, 1\}^{\lvert V\rvert}$, where $x_i = 1$ if $i\in V'$ and $x_i = 0$ otherwise. 
The MIS problem is formulated as
\begin{equation}
    \max_{\mathbf{x}\in \{0,1\}^{|V|}} \sum_{i\in V}x_i \quad
    \text{s.t.} \; x_ix_j \neq 1 \; \text{where} \; (i, j) \in E.
\end{equation}

For this constrained formulation  of MIS, the cost Hamiltonian $C_{\text{MIS}}$ is constructed as $\sum_{i\in V} Z_i$, and the mixing Hamiltonian $B_{\text{MIS}}$ is implemented as a bit flip mixer introduced in Hadfield et al.~\cite{hadfield2019quantum}:
\begin{equation}
\sum_{i \in V} 2^{-\text{deg}(i)} X_{i}
\displaystyle\prod_{k \in \mathcal{N}(i)} (\mathbb{I} \ + Z_k),
\end{equation}
\noindent where $\text{deg}(i)$ is the degree of vertex $i$, and $\mathcal{N}(i)$ denotes the neighborhood of vertex $i$. 
The bit-flip mixer ensures feasible solutions from the QAOA circuit. 
The initial quantum state for the constrained MIS circuit is $\ket{0}^{\otimes n}$.

\subsection{Graph Embedding}

Graph embedding refers to a family of methods that transform vertices, edges, and entire graphs, along with their features, into low-dimensional vector representations.
These embeddings capture key structural and semantic information of the graphs, making them suitable for downstream machine learning tasks.
Representative methods can be categorized into node~\cite{sybrandt2019fobe,ding2020unsupervised,grover2016node2vec} and whole graph embedding~\cite{wang2021graph,cai2018simple,galland2019invariant,narayanan2017graph2vec}.
Node embedding methods focus on mapping individual nodes within a graph to vector representations while preserving local structural information, such as neighborhood connectivity or node attributes. These methods are particularly useful for tasks such as node classification, link prediction, or community detection.
In contrast, whole-graph embedding techniques aim to encode the entire graph into a single fixed-size vector that captures global structural and semantic properties. Such approaches are better suited for graph-level tasks, including graph classification and regression.

In this paper, we design a retrieval model that learns whole-graph embeddings to identify suitable donors for parameter transfer, taking inspiration from information retrieval, such as text-based search~\cite{li2025matchinggenerationsurveygenerative} and recommendation systems~\cite{huang2024comprehensivesurveyretrievalmethods}.
Specifically, we implement a neural network that predicts a transfer score for any donor–acceptor pair, and during inference, we use these predicted scores to retrieve the donor with the highest potential for effective parameter transfer.

%% file: sec/method.tex
\section{Method}

\begin{figure*}
    \centering
    \includegraphics[width=0.55\linewidth]{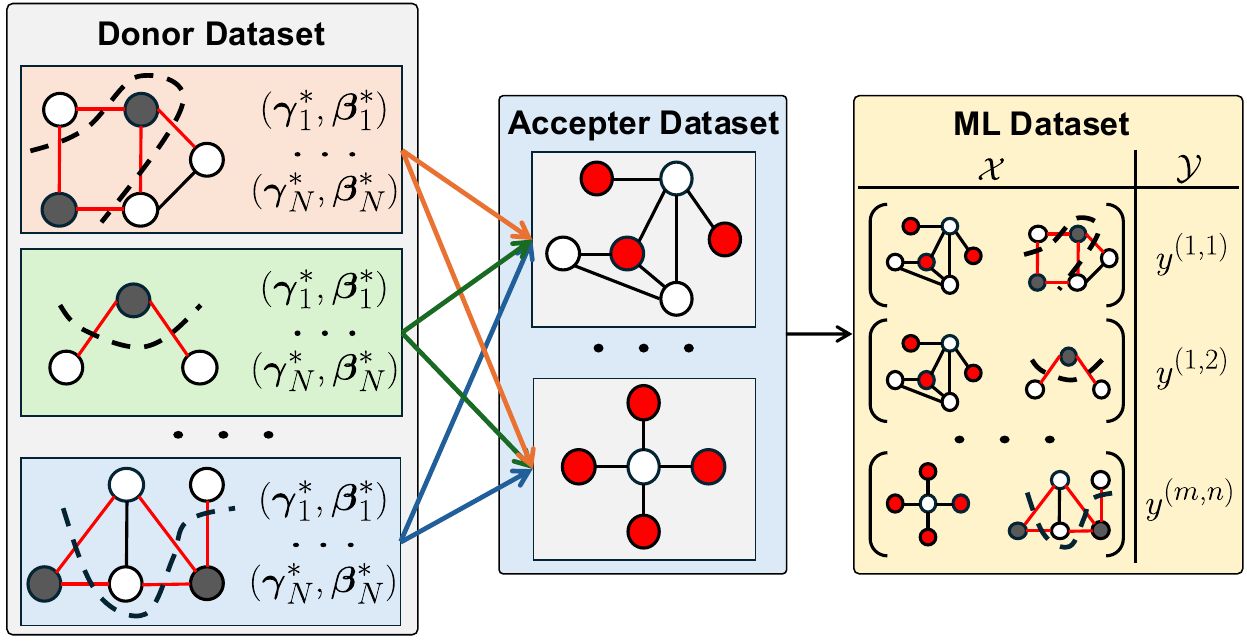}
    \caption{During the data generation procedure, for each instance in the donor dataset $\mathcal{D}$, we train the QAOA circuit with $N$ random initializations $(\boldsymbol{\gamma}_1, \boldsymbol{\beta}_1) \dots (\boldsymbol{\gamma}_N, \boldsymbol{\beta}_N)$ on MaxCut. Then, we directly apply the $N$ optimized parameter sets $(\boldsymbol{\gamma}^*_1, \boldsymbol{\beta}^*_1) \dots (\boldsymbol{\gamma}^*_N, \boldsymbol{\beta}^*_N)$ to each MIS instance in the acceptor dataset $\mathcal{A}$ to retrieve the ground-truth transfer score, which is the average approximation ratios of the optimized MaxCut parameters on MIS.}
    \label{fig:pairs}
\end{figure*}

\subsection{Dataset Generation}
To train the ML model, we first generate a graph dataset $\mathcal{G} = \mathcal{A} \cup \mathcal{D}$ ($\mathcal{A}$ and $\mathcal{D}$ are not necessarily disjoint), where $\mathcal{A}=\{G_{\text{MIS}}^{(i)}\}^{m}_{i=1}$ and $\mathcal{D}=\{G_{\text{MC}}^{(j)}\}^{n}_{j=1}$ are the set of acceptor and donor graphs, respectively.
Then, we employ $N$ random initializations and solve QAOA for the optimized MaxCut parameters $\{(\boldsymbol{\gamma}^*_{t,\text{MC}}, \boldsymbol{\beta}^*_{t,\text{MC}})\}^N_{t=1}$ for each of $n$ instances of $\mathcal{D}$.
We directly apply the $N$ optimized parameter sets from each donor to the acceptor QAOA circuit and record the average expectation value on the acceptor instance as the ground-truth transfer score $y$:
\begin{equation}
    y = \frac{1}{N}\sum^N_{t=1} \bra{\boldsymbol{\gamma}^*_{t,\text{MC}}, \boldsymbol{\beta}^*_{t,\text{MC}}}C_{\text{MIS}}\ket{\boldsymbol{\gamma}^*_{t,\text{MC}}, \boldsymbol{\beta}^*_{t,\text{MC}}}.
\end{equation}
As a result, we construct the ML dataset of acceptor-donor pairs $\mathcal{S} = \{(G_{\text{MIS}}^{(i)}, G_{\text{MC}}^{(j)}, y^{(i,j)})\}$ with $mn$ data points. 
We further split $\mathcal{S}$ into the training set $\mathcal{S}_{\text{train}}$, validation set $\mathcal{S}_{\text{val}}$, and testing set $\mathcal{S}_{\text{test}}$ for the ML training procedure. 

\subsection{Transfer Score Prediction Model}

We construct the score prediction model $f(\cdot, \cdot)$ by training on $\mathcal{S}_{\text{train}} = \{(G_{\text{MIS}}^{(i)}, G_{\text{MC}}^{(j)},y^{(i,j)}) | (i,j) \in \mathcal{I}_{\text{train}}\}$, where $\mathcal{I}_{\text{train}} \subseteq \{1, \dots,m\}\times \{1, \dots, n_\text{train}\}$ and $n_\text{train} = \lvert \mathcal{I}_{\text{train}}\rvert$ is the size of the training set. Specifically, each triple consists of two graph embeddings $G_{\text{MIS}}^{(i)}$ and $G_{\text{MC}}^{(j)}$, and a corresponding label $y^{(i,j)} \in \mathcal{Y} \subseteq [0,1]$ indicating the transfer score of the optimized QAOA parameters from the donor graph to the accepter graph.

To obtain graph embeddings $X \in \mathcal{X} \subset \mathbb{R}^{d_{\text{graph}}}$, where $d_{\text{graph}}$ is the embedding dimension, we employ either a Graph Neural Network (GNN) or a well-established graph embedding method such as Graph2Vec~\cite{narayanan2017graph2vec}, which will be described in more details later in this section. 
In the GNN-based approach, a network $g(\cdot)$ is trained to capture relevant structural properties of the input graph, produce a fixed-dimensional vector embedding, and is trained end-to-end with the score prediction model $f(\cdot,\cdot)$.
Alternatively, methods such as Graph2Vec learn embeddings in an unsupervised manner by maximizing similarity among structurally similar graphs.

\vspace{5pt}
\noindent\textbf{Graph Neural Network.} A standard GNN operates via message passing and pooling. For a graph $G=(V,E)$ with node features $X_{v}$, we initialize each node embedding as $h^{(0)}_v = X_v$. At layer $\ell$, node $v$ updates its embedding $h^{(\ell + 1)}_v$ by aggregating information from its neighbors $\mathcal{N}(v)$.
\begin{equation}
    h^{(\ell + 1)}_v = \text{Update}\bigg(h^{(\ell)}_v, \sum_{u\in\mathcal{N}(v)} \text{Message}\Big(h^{(\ell)}_v,h^{(\ell)}_u,E_{uv}\Big) \bigg),
\end{equation}
where $E_{uv}$ are edge features. After $L$ layers, the node embeddings are pooled with a function $\rho$ (i.e. sum or average) to produce a fixed-dimensional graph embedding $X_G$.
\begin{equation}
    X_G = \rho(\{h^{(L)}_v | v\in V\}).
\end{equation}

\vspace{5pt}
\noindent\textbf{Graph Embedding Method.} Graph embedding methods, such as Graph2Vec~\cite{narayanan2017graph2vec}, convert the entire graph into a fixed-dimensional vector by capturing structural and topological information in an unsupervised manner.
Taking inspirations from natural language processing techniques, Graph2Vec, in particular, treats substructures within a graph as ``words" in a ``document."
It then learns embeddings by maximizing similarities among graphs that share common substructure ``vocabularies."
The approach is task-agnostic and is suitable for downstream tasks such as classification or regression.

In addition to the graph embeddings, we incorporate a trainable, problem-specific context layer that assigns a distinct vector representation to each COP. These vectors capture the unique characteristics of each COP and ensure that the model can effectively distinguish which COP the input graph embeddings belong to.
We first define $\mathcal{P} = \{1,2,\dots,\lvert \mathcal{P}\rvert\}$ as the discrete set of COPs, where each element $p\in\mathcal{P}$ represents a particular COP.
In this case, $\lvert \mathcal{P}\rvert=2$.
The context layer is a learnable matrix $W\in\mathbb{R}^{\lvert \mathcal{P}\rvert\times d_{\text{cop}}}$, where $d_{\text{cop}}$ is the embedding dimension, and each row in $W$ corresponds to a distinct COP $p$.
Lastly, we define an embedding function $e: \mathcal{P} \rightarrow \mathbb{R}^{d_{\text{cop}}}$ such that $e(p) = W_p$; given a COP $p$, its embedding is the corresponding row in $W$.

We then combine the graph embedding $X$ and its associated COP $p$ into a single vector:
\begin{equation}
    z = [X \Vert e(p)] \in \mathbb{R}^{d_{\text{graph}} + d_{\text{cop}}},
\end{equation}


where $[\cdot \Vert \cdot]$ denotes vector concatenation. The combined vectors for the accepter and donor graphs, $z_{\text{MIS}}$ and $z_{\text{MC}}$, are then passed through the score prediction model, which is a fully connected neural network with residual connections~\cite{he2016deep} that subsequently outputs the predicted transfer score $\hat{y}$:
\begin{equation}
    \hat{y} = f(z_{\text{MIS}}, z_{\text{MC}}).
\end{equation}
We use the Mean Squared Loss (MSE) to supervise the model:
\begin{equation}
    \mathcal{L}_{\text{MSE}} (\hat{y},y) = \frac{1}{n_{\text{train}}}\sum_{i=1}^{n_{\text{train}}} (\hat{y}_i - y_i)^2.
\end{equation}
The parameters of the neural networks, $f$ and $g$, and the matrix $W$ are updated via standard gradient descent to iteratively reduce the loss.

\subsection{Donor Retrieval}

During training, the validation set $\mathcal{S}_{\text{val}}$ helps monitor model performance on data not encountered during optimization, guiding the selection of the model that achieves the lowest MSE loss and thus demonstrates the strongest generalization capability. The best model is then employed on the test set $\mathcal{S}_{\text{test}}$ to identify the good donor candidates. Specifically, for each acceptor graph $G^{(i)}_{\text{MIS}} \in \mathcal{S}_{\text{test}}$, the model predicts the transfer score between $G^{(i)}_{\text{MIS}}$ and each donor graph $G^{(j)}_{\text{MaxCut}} \in \mathcal{D}$:
\begin{equation}
    \hat{y}^{(i,j)} = f\Big(g\big(G^{(i)}_{\text{MIS}}\big), g\big(G^{(j)}_{\text{MC}}\big)\Big).
\end{equation}
We achieve a score vector $\boldsymbol{\hat{y}} = (\hat{y}^{(i,1)},\dots, \hat{y}^{(i,n)})$. Then we sort the predicted scores $\boldsymbol{\hat{y}}$ in descending order and select the top-$k$ donor graphs as the candidates for parameter transfer.
In our experiments, $k$ is set to 5.

%% file: sec/experiment.tex
\section{Empirical Results}

\subsection{Dataset Generation Procedure}
\noindent\textbf{Donor Dataset.} We begin by assembling a dataset of 1,000 donor graphs with sizes ranging from 8 to 12 nodes.
Specifically, the dataset includes (1) 250 Erd\"{o}s–R\'{e}nyi graphs with edge probabilities spanning 0.4 to 0.75, (2) 250 random-regular graphs with degrees ranging from 3 to 7 (ensuring that the product of the node count and the degree is even), (3) 250 Watts–Strogatz graphs that have 3 neighbors per node and rewiring probabilities between 0.4 and 0.75, and (4) 250 Barab\'{a}si–Albert graphs where newly added nodes attach between 2 and 4 edges.
For each graph, we optimize the MaxCut QAOA circuit from $N=16$ random initializations using PennyLane~\cite{bergholm2018pennylane} statevector simulation.
We conduct our experiments on QAOA circuits with depths $p=1$ and $p=2$; we use the classical Adam optimizer~\cite{kingma2014adam} to update the parameters.
For each run, we set the maximum number of steps to 200 and the convergence tolerance to 1e-6.

\vspace{5pt}
\noindent\textbf{Acceptor Dataset.}
We use slightly different graph generation rules to construct our acceptor dataset, which contains 500 graphs, also ranging from 8 to 12 nodes.
This is to introduce some structural differences between the donor and acceptor datasets, enabling a more robust evaluation of parameter transfer methods.
Particularly, the dataset includes (1) 200 Erd\"{o}s–R\'{e}nyi graphs with edge probabilities spanning 0.1 to 0.5, (2) 150 random-regular graphs with degrees ranging from 2 to 5 (ensuring that the product of the node count and the degree is even), (3) 50 Watts–Strogatz graphs that have 3 neighbors per node and rewiring probabilities between 0.01 and 0.1, and (4) 100 Barab\'{a}si–Albert graphs where newly added nodes attach between 2 and 4 edges.
We use the Gurobi solver to find the optimal solution for MIS.

\vspace{5pt}
\noindent\textbf{Machine Learning Dataset.}
Combining the donor and acceptor datasets, we construct the machine learning dataset that contains 500 $\times$ 1,000 $=$ 500,000 triples. We further split the number of acceptor graphs into train/validation/test as follows: (1) Erd\"{o}s–R\'{e}nyi: 100/50/50, (2) random-regular: 50/50/50, (3) Watts–Strogatz: 20/10/20, (4) Barab\'{a}si–Albert: 40/20/40, resulting in 210,000 triples for training, 130,000 for validation, and 160 acceptor graphs for testing.

\subsection{Machine Learning Model Implementation}
Our neural network contains two parts: a graph encoder, which is either a graph embedding method (i.e. Graph2Vec) or a Graph Neural Network (GNN) to embed the graphs into fixed-dimensional vectors, and a Fully Connected Network (FCN) as the transfer score prediction model.
We use the PyTorch deep learning framework to implement the neural network.
We train the models for 20 epochs using the AdamW optimizer~\cite{loshchilov2017decoupled} with a batch size of 256.
We set the initial learning rate to 1e-4 and apply the reduce-on-plateau scheduler which reduces the learning rate by a factor of 0.05 when the model's performance stops improving on the validation set.

\vspace{5pt}
\noindent\textbf{Graph2Vec Implementation.} To generate the graph embeddings, we use KarateClub's implementations~\cite{rozemberczki2020karate}. Using the default settings, we set the number of epochs to 10, the learning rate to 0.025, the number of Weisfeiler-Lehman iterations to 2, and the downsampling frequency to 1e-4.
The dimensionality  of the embeddings is set to $d_{\text{graph}}=128$.
We directly use Graph2Vec (G2V) embeddings as input to our score prediction model without any modifications.
Furthermore, following Falla et al.~\cite{falla2024graph}, we examine whether parameter transferability could be driven by the similarity of the Graph2Vec embeddings.
In particular, we pick the $k$ donor graphs whose Graph2Vec embeddings are closest (in Euclidean distance) to that of the acceptor graph.
We dub this method ``Closeness" in Tab.~\ref{tab:transfer}.

\vspace{5pt}
\noindent\textbf{Graph Neural Network Implementation.} We experiment with different GNN architectures, namely Graph Convolutional Network (GCN)~\cite{kipf2016semi}, Graph Attention Network (GAT)~\cite{velivckovic2017graph}, Higher-order GNN (GraphConv)~\cite{morris2019weisfeiler}, and Chebyshev Convolutional Network (ChebConv)~\cite{defferrard2016convolutional}.
We implement a 5-block GNN with each neural network block having the following architecture: GNN layer $\rightarrow$ BatchNorm~\cite{ioffe2015batch} $\rightarrow$ ReLU $\rightarrow$ Dropout~\cite{srivastava2014dropout}.
We set the Dropout rate to 0.2 during training.
We provide six node-level features as input to the GNN: node degree, clustering coefficient, \(k\)-core number, betweenness centrality, PageRank score, and the count of triangles to which the node belongs.

\vspace{5pt}
\noindent\textbf{Score Prediction Model Implementation.} The score prediction model contains 4 fully connected blocks with residual connections~\cite{he2016deep}. The architecture of each block is as follows: Linear $\rightarrow$ LayerNorm~\cite{ba2016layer} $\rightarrow$ ReLU.
It is then followed by a Linear layer that outputs a scalar, which is the predicted transfer score $\hat{y}$.

\begin{table*}[t!]
\caption{Performance in terms of mean best approximation ratio and mean best probability of finding the optimal solution.}
\label{tab:transfer}
\centering
    \resizebox{0.75\linewidth}{!}{
    \begin{tabular}{|l|c|c|c|c|}
    \hline
    Circuit Depth & \multicolumn{2}{c|}{$p=1$} & \multicolumn{2}{c|}{$p=2$} \\
    \hline
    Method & Mean Best $r$ & Mean Best $\text{probability}(^*x)$ & Mean Best $r$ & Mean Best $\text{probability}(^*x)$ \\
    \hline
    GCN       & 0.8102           & 0.4718                   & 0.8261           & 0.4674                   \\
    \hline
GAT       & 0.8018           & 0.4452                   & 0.8142           & 0.4473                   \\
\hline
GraphConv & 0.8192           & 0.4927                   & 0.8293           & 0.4816                   \\
\hline
ChebConv  & 0.7999           & 0.4436                   & 0.8370           & 0.4889                   \\
\hline
G2V       & 0.8035           & 0.4536                   & 0.8385           & 0.4958                   \\
\hline
Closeness & 0.8072           & 0.4641                   & 0.8062           & 0.4216                  \\    
    \hline
    \end{tabular}
}

\end{table*}

\begin{figure}
    \centering
    \includegraphics[width=0.9\linewidth]{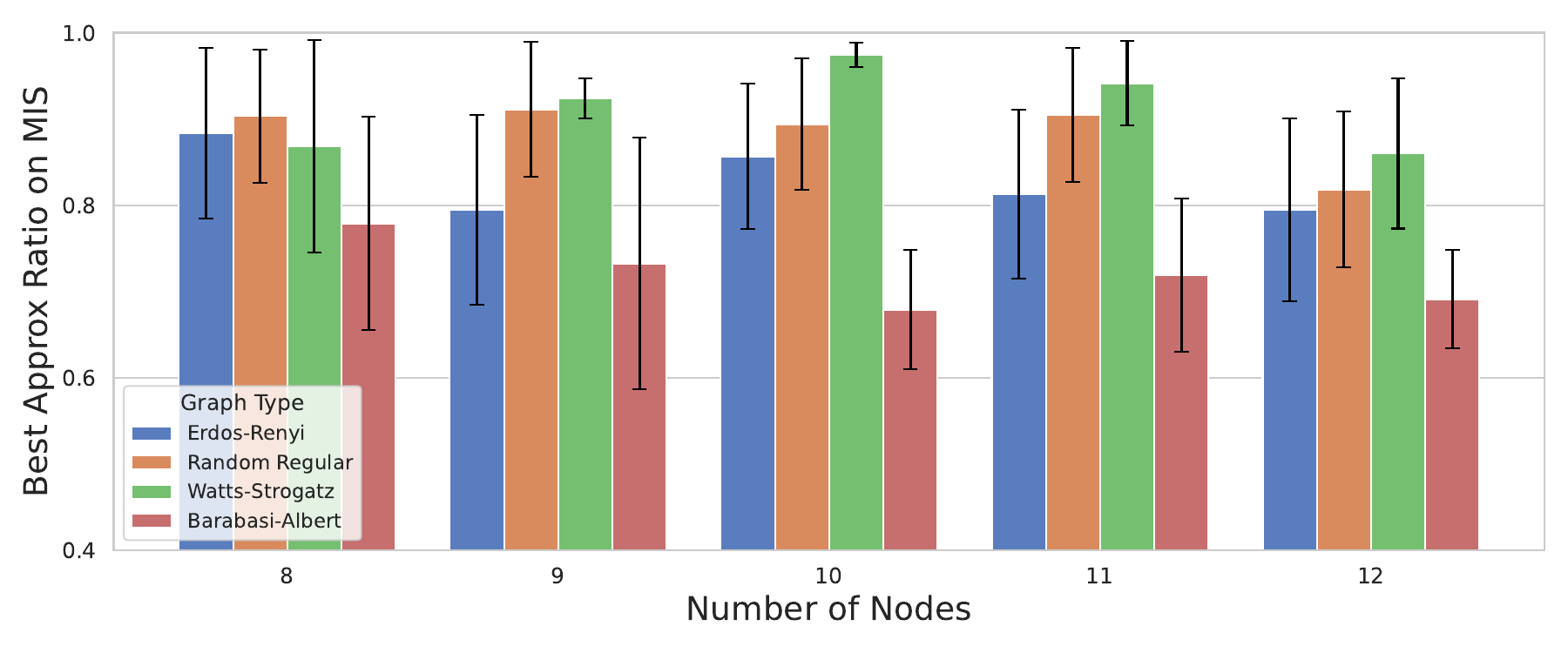}
    \caption{Performance on the approximation ratio $r$ achieved by the best transferred MaxCut parameters from our ChebConv-based score prediction model for $p=2$, shown across different graph types and number of nodes.}
    \label{fig:transfer}
\end{figure}

\begin{figure}
    \centering
    \includegraphics[width=0.9\linewidth]{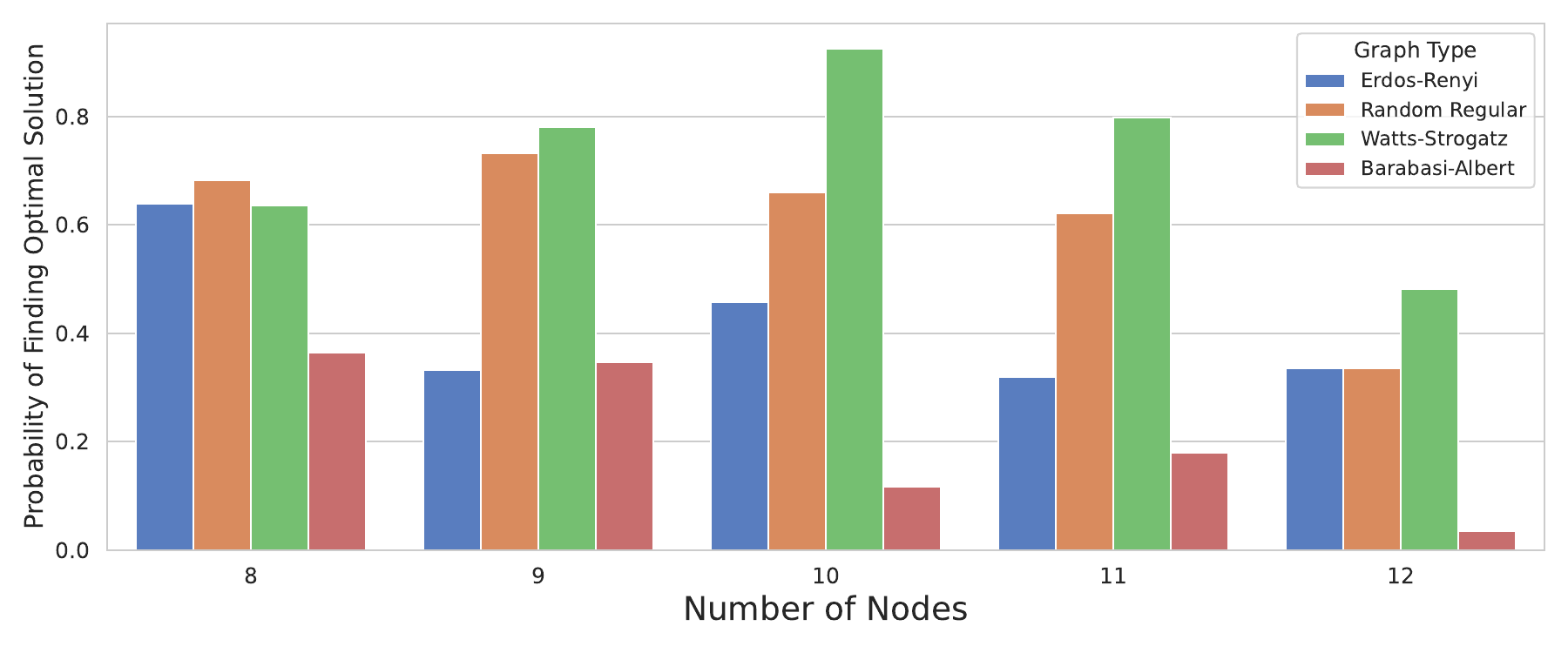}
    \caption{Performance on the probability of finding the optimal solution $\text{probability}(^*x)$ achieved by the best transferred MaxCut parameters from our ChebConv-based score prediction model for $p=2$, shown across different graph types and number of nodes.}
    \label{fig:prob-hit}
\end{figure}

\subsection{Results on Parameter Transferability}

We evaluate the quality of the donor graphs identified by our machine learning model.
We pick $k=$ 5 following the standard practice of the instance retrieval research in machine learning~\cite{radford2021learning}.
Once the top-5 donors are chosen for each acceptor graph in the test set $\mathcal{S}_{\text{test}}$, we take each donor's 16 optimized MaxCut parameters, apply them to the MIS circuit, and perform QAOA inference for 1,000 shots to measure the approximation ratio $r$ and the probability of finding the optimal solution $\text{probability}(^*x)$.
We record the best-performing parameter set among the five donors based on the metrics and present the results in Tab.~\ref{tab:transfer}.

Regarding the mean best approximation ratio $r$ for $p=2$, ChebConv and Graph2Vec achieve the best results with 0.8370 and 0.8385.
The Closeness approach delivers fair results but generally is still the lowest among all methods, highlighting the importance of further fine-tuning the original Graph2Vec embeddings to harness the full potential.
We further illustrate the results for ChebConv by graph types and number of nodes in Fig.~\ref{fig:transfer}.
We see that Barab\'{a}si–Albert graphs achieve the lowest overall approximation ratio, and the mean best approximation ratio gradually decreases as the node count grows, signifying the increased difficulty of transferring parameters to larger graphs.
A similar pattern is observed in the probability of finding the optimal solution, indicating that Barab\'{a}si–Albert to be a particularly challenging class of graph for this setting.


\begin{figure}
    \centering
    \includegraphics[width=0.73\linewidth]{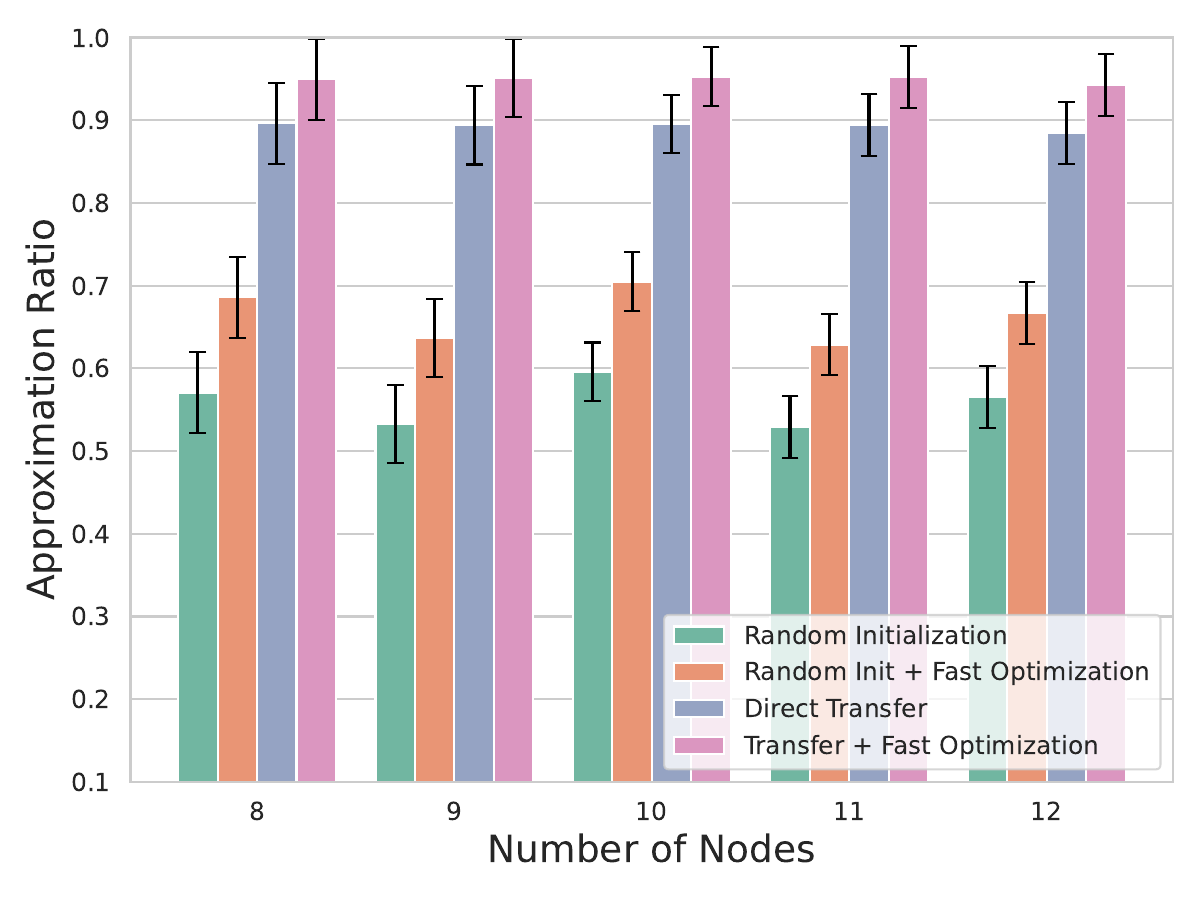}
    \caption{For $p=2$, using the best transferred MaxCut parameters with $r \geq$ 0.8 as initializations for MIS, we further optimize the QAOA circuit for just 10 additional steps. This brief post-optimization boosts the average approximation ratio from 0.8933 to 0.9500. The results are from the ChebConv-based model.}
    \label{fig:post-optim}
\end{figure}

\subsection{Results on Optimization Speedup}

In this subsection, we show how leveraging the transferred MaxCut parameters can significantly accelerate the optimization process and yield higher approximation ratios compared to starting from random initializations.
We first select the best transferred parameters that yield an approximation ratio $r \geq$ 0.8, use them as intializations for the MIS circuit, and perform QAOA optimization for \textit{only 10 steps}.
This brief post-optimization phase significantly boosts the average approximation ratio to 0.9500.
In addition, we plot our result against the average approximation ratio of direct transfer of MaxCut parameters without further optimization, random initialization of MIS parameters and random initialization of MIS parameters with 10-step optimization in Fig.~\ref{fig:post-optim}.
For comparison, we also optimize each MIS circuit from 16 random initializations and record that it takes 155 steps to reach an average approximation ratio of 0.8833.
These results underscore the gains in efficiency and performance of parameter transfer, demonstrating how reusing carefully chosen MaxCut parameters substantially cuts down on optimization time while producing superior MIS solutions compared to random initializations.

%% file: sec/conclusion.tex
\section{Conclusion and Future Outlook}
In this work, we investigated the transferability of QAOA parameters from one unconstrained combinatorial problem (MaxCut) to another constrained one (Maximum Independent Set). {\it This is an important step towards building comprehensive AI model for quantum circuit generation for novel problems.}
By adopting a data-driven approach, we demonstrated how a machine learning model, utilizing both graph embeddings and a problem-specific context layer, can systematically identify good donor candidates {\it from another problem}.
Our empirical studies have shown that the transferred parameters have significantly reduced optimization time and improved approximation ratios compared to starting from random initializations on MIS.
For future research, we aim to extend our study to deeper circuit depths, larger graphs (which is more challenging due to problems such as barren plateaus), and multiple combinatorial optimization problems. This will require substantially more extensive computational resources.

\section*{Acknowledgements}
This research is supported by NSF award \# 2427042.
